\documentclass{elsart}

\usepackage{epsfig}

\begin{document}

\begin{frontmatter}
  
\title{\large \bf Search for Lepton Flavor Violation Process
    $J/\psi \to e \mu$}

\date{5 Mar. 2003}

\maketitle

\begin{center}

J.~Z.~Bai$^1$,        Y.~Ban$^{9}$,          J.~G.~Bian$^1$,
       X.~Cai$^{1}$,          J.~F.~Chang$^1$,
H.~F.~Chen$^{16}$,    H.~S.~Chen$^1$,
J.~Chen$^{3}$,        Jie~Chen$^{8}$,        J.~C.~Chen$^1$
% ...
\vspace{0.2cm}
\\(BES Collaboration)\\

\vspace{0.2cm}

$^1$ Institute of High Energy Physics, Beijing 100039, People's Republic of
     China\\
% ...
$^3$ Colorado State University, Fort Collins, Colorado 80523\\
% ...
$^8$ Nankai University, Tianjin 300071, People's Republic of China\\
$^{9}$ Peking University, Beijing 100871, People's Republic of China\\
% ...
$^{16}$ University of Science and Technology of China, Hefei 230026,
        People's Republic of China\\
% ...

\vspace{0.4cm}

\end{center}

\normalsize

\begin{abstract}
  {Using a sample of $5.8 \times 10^7 J/\psi$ events, the Beijing Spectrometer
    experiment has searched for the decay $J/\psi \to e \mu$. Four
    candidates, consistent with the estimated background, are
    observed, and an upper limit on the branching fraction of $J/\psi
    \to e \mu$ of $1.1 \times 10^{-6}$ at the 90 $\%$ C.L. is
    obtained.}
\vspace{3\parskip}
%\noindent{\it PACS:} 13.25.Gv, 14.40.Gx, 13.40.Hq
\end{abstract}
\end{frontmatter} 

\clearpage
       
\section{Introduction} 
%%%%%%%%%%%%%%%%%%%%%%%%%%  Following needs work
 In the minimal standard model of electroweak theory,
% the total and individual lepton numbers are conserved, but
the three separate lepton numbers, electron number, muon number, and
tau number, are conserved, but this conservation law 
may be broken in many extensions of the Standard Model, such
 as grand unified models \cite{ref1}, 
supersymmetric standard models \cite{ref2}, left-right symmetric
models \cite{ref3}  and models 
 where electroweak symmetry is broken dynamically \cite{ref4}.
 Recent experimental results from Super-Kamiokande \cite{ref5}, 
 SNO  \cite{ref6} and KamLAND~\cite{kamland}
 indicate strongly
 that neutrinos have masses and mix with each other; hence lepton
 flavor symmetries are broken. There have been  
 many studies both experimentally and theoretically on searching for
 the lepton flavor violating processes \cite{ref7}, mainly 
 from $\mu$, $\tau$ and Z decays~\cite{PDG}. Theoretical studies on the possibility of searching for the lepton flavor violation
 in decays of charmonium and bottomonium systems are discussed in
Refs. \cite{ref8}-\cite{ref10}.
 In this paper we report on a 
search for lepton flavor violation  via the process 
 J/$\psi \to e^\pm\mu^\mp$ using $5.8 \times 10^7 J/\psi$ events at BEPC/BESII .
 
%%%%%%%%%%%%%%%%%%%%%%%%%

\section{BES detector}
The BEijing Spectrometer (BES) \cite{ref11} is a large general purpose
solenoidal detector at the Beijing Electron Positron Collider (BEPC).
The beam energy of BEPC ranges from 1.0 to 2.5 GeV, and the peak
luminosity at the $J/\psi$ is around $5 \times 10^{30}$
cm$^{-2}$s$^{-1}$. The upgrade from BESI to BESII \cite{ref12}
includes the replacement of the central drift chamber with a
straw-tube vertex chamber, composed of 12 tracking layers arranged
around a beryllium beam pipe and with a spatial resolution of about 90
$\mu$m; a new barrel time-of-flight (TOF) counter with a time
resolution of 180 ps; and a new main drift chamber (MDC), which has 40
tracking layers with a $dE/dx$ resolution of $\sigma_{dE/dx} = 8.0 \%$
and a momentum resolution of $\sigma_{p}/p = 1.78\%\sqrt{1+p^2}$ ($p$
in GeV/$c$) for charged tracks. These upgrades augment the
pre-existing calorimeter and muon tracking systems.  The barrel shower
counter (BSC), which has a total thickness of 12 radiation lengths and
covers 80 $\%$ of 4$\pi$ solid angle, has an energy resolution of
$\sigma_E/E=0.21/\sqrt{E}$ ($E$ in GeV) and a spatial resolution of
7.9 mrad in $\phi$ and 2.3 cm in z.  The $\mu$ identification system
consists of three double layers of proportional tubes interspersed in
the iron flux return of the magnet.  They provide coordinate
measurements along the muon trajectories with resolutions in the
outermost layer of 10 cm and 12 cm in $r \phi$ and $z$, respectively.
The absorber thicknesses in front of the three layers are 120, 140,
and 140 mm, and the solid angle coverage of the layers is 67 \%, 67
\%, and 63 \% of $4\pi$, respectively.

\section{Event selection} 
\begin{figure}[htbp]
%\centerline{
%\psfig{figure=sacolemunl.eps,width=7.5cm,height=5.5cm,angle=0}
\psfig{figure=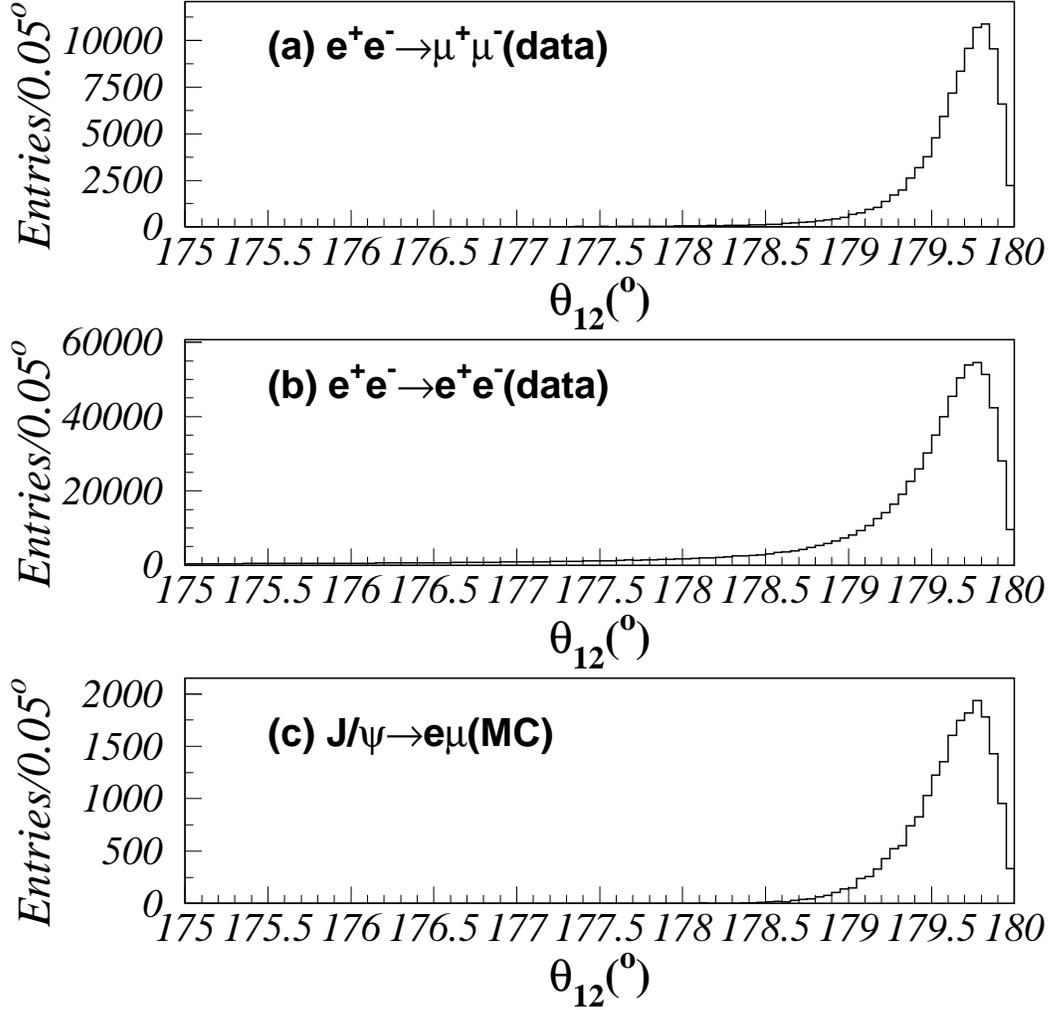,width=15cm,angle=0}
\caption{The opening angle between two charged tracks for
(a) $e^+e^- \to \mu^+ \mu^-$
(data), (b) $e^+e^- \to e^+ e^-$ (data) and (c) $J/\psi \to e \mu$
(Monte Carlo data). }
\label{figure:1}
\end{figure}

The initial selection of $J/\psi \to e^{\pm} \mu^{\mp}$ requires that
candidate events have two oppositely charged tracks within the polar
angle region, $|\cos{\theta}|<0.8$, and fewer than four neutral
tracks.  A charged track should have a good helix fit; a momentum,
$P$, with 1.45 GeV/$c$ $< P < 1.65$ GeV/$c$; and originate from the
interaction region, defined by $|x|<0.015$ m, $|y|<0.015$ m and
$|z|<0.15$ m. To reject cosmic rays, the TOF time difference of the
two charged tracks must be less than 1.2 ns.  The invariant mass of
the $e \mu$ system, $M_{e \mu}$, must satisfy 2.95 GeV/$c^2 <~ M_{e
  \mu}~ <$ 3.25 GeV/$c^2$, and the angle between the two charged
tracks, $\theta_{12}$, is required to be greater than 178.5$^{\circ}$.
The $\theta_{12}$ distribution is shown in Fig.~\ref{figure:1}.
%for $e^+e^- \to \mu^+ \mu^-$ (data), $e^+e^- \to e^+ e^-$ (data), and
%$J/\psi \to e \mu$ (Monte Carlo data).

Isolated photons are defined as those photons having an 
energy deposit in the BSC greater than 50 MeV, an angle
with any charged track greater than $15^{\circ}$,
and an angle between the direction defined by the first layer hit in
the BSC
and the developing direction of the cluster in the $xy$-plane less than
$18^{\circ}$.  There must be no isolated photon in the selected event.

\begin{figure}[htbp]
\psfig{figure=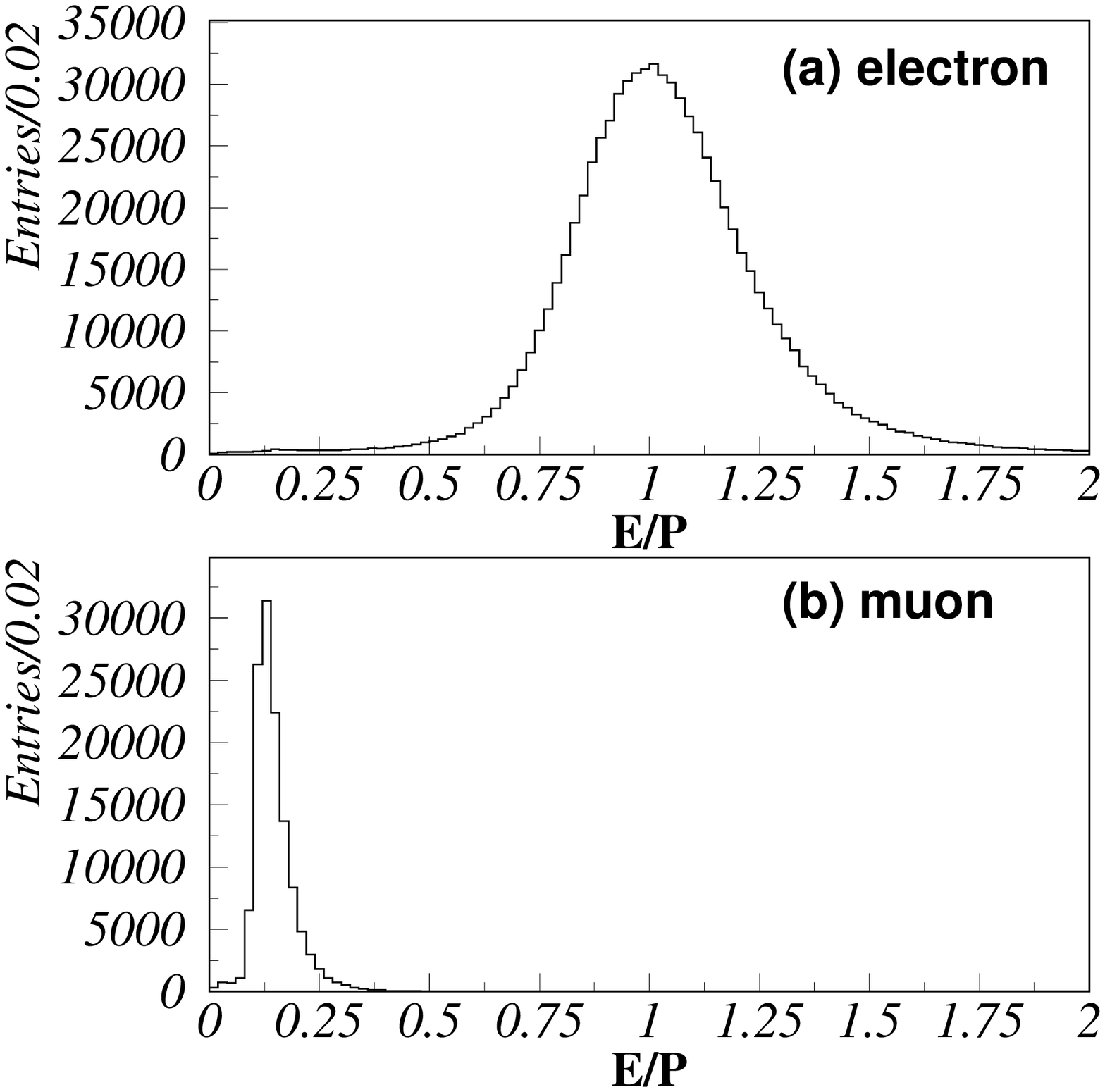,width=15cm,angle=0}
\caption{Distribution of
  $E/P$ of candidate a.) $e$ and b.) $\mu$ tracks. $E$ and $P$ denote
  the energy deposited in the BSC and the momentum measured in the
  MDC, respectively.}
\label{figure:2}
\end{figure}

%%%%%%%%%%%%%%%%%%%%%%Put back the way it was
The above criteria select back-to-back, two prong events, such as
$J/\psi \to e^+ e^-(\gamma), \mu^+ \mu^-(\gamma), K^+ K^-, \pi^+
\pi^-$ and $e^+ e^- \to e^+ e^- (\gamma)$, etc. To select electrons and
muons, the BSC and $\mu$ counter information is used.
The actual selection criteria are based on distributions determined
from data.
Fig.~\ref{figure:2} shows distributions of $E/P$, where $E$ is the energy
deposited in
the shower counter, for candidate electron and muon tracks.
To be an electron, a track must have no hits in the muon counter
and satisfy  $E/P > 0.7$.
  
\begin{figure}[htbp]
%\centerline{
%\psfig{figure=sample_mu3l.eps,width=7.5cm,height=5.5cm,angle=0}
\psfig{figure=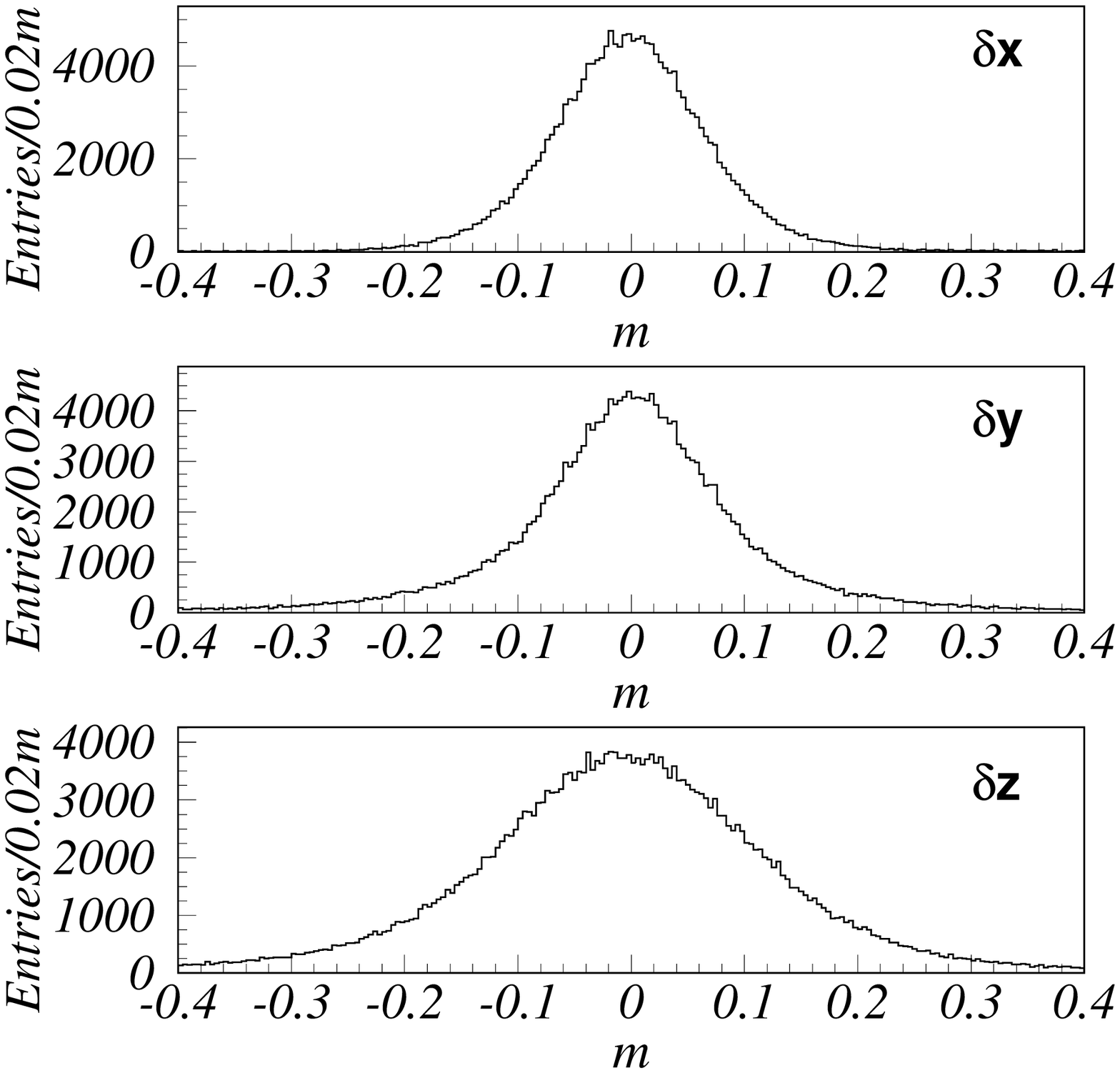,width=15cm,angle=0}
\caption{The differences, $\delta x$, $\delta y$, and $\delta z$, between the closest hit and the extrapolated
  MDC track in the third layer of the $\mu$ system.}
\label{muhits}
\end{figure}

To select muon tracks, the differences, $\delta_i (i = x, y, z)$,
between the closest muon hit and the projected MDC track in each layer
are used.  Fig.~\ref{muhits} shows these differences in the third
layer of the $\mu$ counter.  The distributions are Gaussians, and
standard deviations, $\sigma_i (i = x, y, z)$, are determined for each
layer of the $\mu$ counter. 

 To reduce background, we make a tight cuts
on the muon. A good hit in the $\mu$ counter requires
$|\delta_i| < \sigma_i$ for $i$ = x, y, and z.  The total number of
layers hit in the $\mu$ counter is denoted as $\mu_{hit}$ and can
range from zero to three.  The number of good hits is denoted by
$\mu_{hit}^{good}$.
A track is considered as a muon if $E/P < 0.3$ and
$\mu_{hit}^{good}$ is
  greater than zero if the transverse momentum
of the track, $P_{xy}$, is less than 0.75 GeV/$c$, $\mu_{hit}^{good} >1$ if
0.75 GeV/$c$ $< P_{xy} < 0.95$ GeV/$c$, or $\mu_{hit}^{good}=3$ if
$P_{xy} > 0.95$ GeV/$c$. 

%Events with at least one track satisfy the electron criteria is 
%considered as Bhabha events, and for reasons to be discussed later, 
%the ratio of two electron to one electron events gives the 
%electron selection efficiency, as listed in Table 3. Similarly, 
%events with at least one track satisfy the muon criteria is 
%considered as di-muon events, and the ratio of two muon to one muon 
%events gives the muon selection efficiency, again listed in Table 3. 

Using these selection criteria, 
four $J/\psi \to e \mu$ candidates are found  
in 
$5.8 \times 10^7 J/\psi$ events.  The characteristics of these events are
listed in Table~\ref{table3}.

%%%%%%%%%%%%%%%%%%%%%%%%%%%table needs work 
\begin{table}[htb]
\begin{center}
\caption{The characteristics of the $J/\psi \to e \mu$ candidates.
$\theta_{12}$ is the angle between two charged tracks.
\label{table3}
}
%\begin{tabular}{lcccc} \hline
% & RUN:14676 & RUN:16419 & RUN:18940 & RUN:19149 \\
% & REC:16877 & REC:26352 & REC:27010 & REC:9196 \\ \hline
% $P$ (GeV/$c$)& 1.591 | 1.522 & 1.570 | 1.543 & 1.587 | 1.611 &
% 1.605 | 1.535 \\
% $E/P$ & 0.7188 | 0.1758 & 0.8577 | 0.1778 & 0.8835 | 0.1415 &
% 0.7155 | 0.1035 \\
% $\mu_{hit}$ & 0 | 3 & 0 | 3 & 0 | 3 & 0 | 3  \\
% $M_{e \mu}$ (GeV) & 3.113 & 3.117 & 3.201 & 3.143  \\
% $\theta_{12}$ & 179.5 & 179.8 & 179.6 & 179.5  \\ \hline
%\end{tabular}
%\end{center}
%\end{table}

\begin{tabular}{l|cc|cc|cc|cc} \hline
RUN No. & \multicolumn{2}{c|}{14676} & \multicolumn{2}{c|}{16419} & \multicolumn{2}{c|}{18940} & \multicolumn{2}{c}{19149} \\
REC No. & \multicolumn{2}{c|}{16877} & \multicolumn{2}{c|}{26352} & \multicolumn{2}{c|}{27010} & \multicolumn{2}{c}{9196} \\ \hline
 $M_{e \mu}$ (GeV) & \multicolumn{2}{c|}{3.113} & \multicolumn{2}{c|}{3.117} & \multicolumn{2}{c|}{3.201} & \multicolumn{2}{c}{3.143}  \\
 $\theta_{12}$ & \multicolumn{2}{c|}{179.5} & \multicolumn{2}{c|}{179.8} & \multicolumn{2}{c|}{179.6} & \multicolumn{2}{c}{179.5}  \\ \hline
track  & $e$ & $\mu$ & $e$ & $\mu$ & $e$ & $\mu$ & $e$ & $\mu$ \\
 $P$ (GeV/$c$)& 1.591 & 1.522 & 1.570 & 1.543 & 1.587 & 1.611 &
 1.605 & 1.535 \\
 $E/P$ & 0.7188 & 0.1758 & 0.8577 & 0.1778 & 0.8835 & 0.1415 &
 0.7155 & 0.1035 \\
 $\mu_{hit}^{good}$ & 0 & 3 & 0 & 3 & 0 & 3 & 0 & 3  \\ \hline
\end{tabular}
\end{center}
\end{table}

\section{Efficiencies and backgrounds}

The main backgrounds for $J/\psi \to e \mu$ are nearly back-to-back
$J/\psi \to e^+ e^-(\gamma)$, $\mu^+ \mu^-(\gamma)$, $K^+ K^-$, $\pi^+
\pi^-$, $e^+ e^- \to e^+ e^- (\gamma)$, and $e^+ e^- \to \mu^+ \mu^-
(\gamma)$ events where one or more tracks is misidentified as a $\mu$
or $e$.  It is therefore important to determine the $\mu$ and $e$
particle identification (PID) misidentification efficiencies for the
background channels, as well as the PID efficiencies for the
signal channel.  In this study, we determine these efficiencies using
information from the BSC and muon counters from data.  The overall
efficiencies include the PID efficiencies and the geometric
efficiencies, $\epsilon_{MC}$, determined using our Monte Carlo (MC)
program SIMBES (SIMulation at BES), which is based on
GEANT3.21.  For the signal channel $J/\psi \to e \mu$,
the total selection efficiency is then
$$\epsilon_{J/\psi \to e \mu} = \epsilon_{e \mu-MC} \times \epsilon_{e
  \to e} \times \epsilon_{\mu \to \mu},$$
where $\epsilon_{e \mu-MC}$
is the geometric efficiency, $\epsilon_{e \to e}$ is the electron PID
efficiency, and $\epsilon_{\mu \to \mu}$ is the muon PID efficiency.
Using 30000 $J/\psi \to e \mu$ Monte Carlo events with an
angular distribution of $1+\cos^2{\theta}$, the geometric
efficiency is determined to be $\epsilon_{e \mu-MC} = (53.7 \pm
0.3)\%$.

For the background channels $J/\psi$ $(e^+e^-) \to XX$, where $X=e$,
$\mu$, $\pi$, $K$, the efficiency after $J/\psi \to e \mu$ selection
is
$$\epsilon_{XX} = \epsilon_{XX-MC} \times \epsilon_{X \to e} \times
\epsilon_{X \to \mu}\times 2,$$
where $\epsilon_{XX-MC}$ is the Monte
Carlo geometric acceptance, shown in Table~\ref{table.7}, and $\epsilon_{X \to
  e}$ and $\epsilon_{X \to \mu}$ are the particle misidentification
efficiencies for X being identified as an
electron or a muon, respectively.

\begin{table}[htbp]
\begin{center}

\caption{Monte Carlo geometric efficiencies. 
\label{table.7}
}
\begin{tabular}{lll}\hline
channel & & MC Efficiency \\\hline
$J/\psi \to ee$  & $\epsilon_{ee-MC}$ & (61.47$\pm$0.02)$\%$\\
$J/\psi \to \mu\mu$ & $\epsilon_{\mu\mu-MC}$ & (58.32$\pm$0.02)$\%$\\
$J/\psi \to \pi\pi$  &$\epsilon_{\pi\pi-MC}$ & (52.74$\pm$0.29)$\%$\\
$J/\psi \to KK$ & $\epsilon_{KK-MC}$ &(24.38$\pm$0.24)$\%$\\
 $e^+e^- \to e^+e^-(\gamma)$  & $\epsilon_{ee(\gamma)-MC}$ &(32.51$\pm$0.03)$\%$\\
 $e^+e^- \to \mu^+ \mu^-(\gamma)$  & $\epsilon_{\mu \mu(\gamma)-MC}$ &(42.96$\pm$0.29)$\%$\\ \hline
     \end{tabular}
\end{center}
\end{table}

The $e$ and $\mu$ PID efficiencies are determined using $e$-pair
samples of $e^+ e^- \to e^+ e^-
(\gamma)$ (Bhabha, resonance, and continuum production)
events and $\mu$-pair samples of $e^+ e^- \to \mu^+ \mu^- (\gamma)$
(resonance and continuum production) events. The highest momentum
track is required to satisfy $1.45 < P < 1.65$ GeV/c and
to be identified as an $e$ or $\mu$ according to the PID selection
criteria.  The other track in the event is then assumed to be the same
type of lepton, and the fraction of these satisfying the lepton
selection determines the efficiency. The PID efficiencies are shown in
Table~\ref{table.5}, and the total efficiency for $J/\psi \to e \mu$ is
$(9.7 \pm 0.5)\%$.

The misidentification efficiencies between electrons and muons cannot
be studied from the data because the events selected for this study
are the same as our $J/\psi \to e \mu$ candidates. 
A rough estimate of this background using a
Monte
Carlo simulation predicts a total of 7 background
events from $e^+ e^- \to e^+e^-(\gamma)$ and $e^+ e^- \to \mu^+
\mu^-(\gamma)$. Since the Monte Carlo simulation is not precise enough
to determine this background, we conservatively
ignore these backgrounds in the estimation of the upper limit on the
branching ratio for
$J/\psi \to e\mu$.

For the misidentification efficiencies of hadrons, $\pi$ and $K$
tracks are selected from $J/\psi \to \rho^{\pm} \pi^{\mp}$ and
$K^{*\pm}K^{\mp}$ decays.  The track with the highest momentum is
considered as the $\pi$ and $K$ in the respective samples, and a total
of 60551 $\pi$ tracks and 9275 $K$ tracks are selected.  The fraction of the
tracks above 1.2 GeV/c passing the electron and muon selection
criteria are used to obtain the misidentification efficiencies, that
are listed in Table~\ref{table.5}.  The resultant background rates from
particle
misidentification of $J/\psi \to K^+ K^-, \pi^+ \pi^-$ events are listed in
Table~\ref{table.6}, taking into account their branching ratios from
PDG2002~\cite{PDG}.

\begin{table}[htbp]
\begin{center}
\renewcommand{\tabcolsep}{0.3pc} % enlarge column spacing
\caption{The particle identification/misidentification efficiencies. \label{table.5}
}
%\vskip -0.3cm
    \begin{tabular}{lll}\hline
                          &regarded as $e$         &regarded as $\mu$ \\\hline
         $e$ sample &95.3$\%$(1$\pm$0.02$\%$) &\hspace{0.5cm} | \\
         $\mu$ sample    &\hspace{0.5cm} |  &19.0$\%$(1$\pm$0.6$\%$)   \\
         $\pi$         &3.6$\%$(1$\pm$2.1$\%$)
    &0.46$\%$(1$\pm$5.98$\%$)\\
       $K$         &3.11$\%$(1$\pm$5.79$\%$) &0.38$\%$(1$\pm$16.8$\%$)\\\hline
      \end{tabular}
%    \end{center}
\end{center}
\end{table} 

\begin{table}[htbp]
\begin{center}
\caption{The misidentification rates and backgrounds from 
hadronic channels. \label{table.6}
}
%\vskip -0.3cm    
\begin{tabular}{lll}\hline
 decays  &misidentification &number of\\
 &rate &  background\\\hline
 J/$\psi \to \pi\pi$ &$1.74 \times 10^{-4}$ &1.49\\
 J/$\psi \to KK$ &$5.77 \times 10^{-5}$ &0.79 \\ \hline
total && 2.3 \\\hline
     \end{tabular}
\end{center}
\end{table}

There are a number of sources of systematic error, including the error
on the total number of $J/\psi$ ($4.7\%$) events, the errors on the
branching ratios from PDG2002~\cite{PDG}, the uncertainties of the
muon and electron identification efficiencies, shown in
Table~\ref{table.5}, and the errors on the hadronic misidentification
efficiencies, also shown in Table~\ref{table.5}.
%%%%%%%%%%%%%%%%%%%%%%%%%%%%%%%%%%%%%
Table~\ref{table.8} summarizes the errors in the background from the hadronic
channels. 
The total background,  ignoring the contributions from  
$J/\psi \to e^+ e^-(\gamma)$ and $\mu^+ \mu^-(\gamma)$,
is $2.3 \pm 0.3$ events.
Therefore the observed four events are consistent with background.

\begin{table}[htbp]
\begin{center}
\caption{Summary of estimated errors for hadronic background
channels. \label{table.8}
}
 \begin{tabular}{lllllll}\hline
        &$N_{total}$ &BR.   & 
 $e_{\rm PID}$ & $\mu_{\rm PID}$    &total error  &number \\\hline
$J/\psi \to \pi\pi$ &4.7$\%$ &15.6$\%$ & 2.10$\%$ & 5.98$\%$
 &17.4$\%$&1.49$\pm$0.26\\
$J/\psi \to KK$ &4.7$\%$ &13.1$\%$ &5.79$\%$ & 16.8$\%$ &22.5$\%$
 &0.79$\pm$0.18\\ \hline
  \end{tabular}
\end{center}
\end{table}

\section{Results and discussion}
The four observed $J/\psi \to e \mu$ candidates
are consistent with the estimated background,
hence an upper limit for $J/\psi \to e \mu$ is determined.
Only the backgrounds from the hadronic channels are taken into account in
the estimation of the upper limit. 

There are several ways to determine the upper limit
 \cite{ref14} \cite{ref15} \cite{ref16}. The
following formula is used to calculate the upper limit
for $Br(J/\psi \to e \mu)$:
$$Br < \lambda(N_{OB},N_{BG})/[N_{T}\times\epsilon_{J/\psi \to e\mu}],$$
where $\lambda$ is the upper bound of a $90\%$ C.L.
 for an unknown Poisson signal mean, for total observed events,
$N_{OB}$, and known mean background, $N_{BG}$.
$N_T$ is the total number of $J/\psi$ events, and
$\epsilon$ is the detection efficiency. $\lambda(N_{OB},N_{BG})$
can be calculated using the method of Ref.~\cite{ref14}.    

As a conservative estimation, we take the number of background events as           
the central value of the number of background events reduced by one standard              
deviation, $\it{i.e.}$ $N_{BG}=2.0$ and $N_{OB}$=4,
then $\lambda(N_{OB},N_{BG})$=5.98.
Therefore we obtain:                                                                 
$$Br(J/\psi \to e \mu)<1.1 \times 10^{-6}.$$   

 In summary, we searched for $J/\psi \to e \mu$ based on $5.8 \times 10^{7} 
J/\psi$ events and observed four $J/\psi \to e \mu$ candidates passing our
selection criteria, which
are consistent with the estimated background. The upper limit of 
$Br(J/\psi \to e \mu)$ is determined to be $1.1 \times 10^{-6}$ at the 90 $\%$ C.L.

\section{Acknowledgments}
 The BES collaboration thanks the staff of BEPC for their hard efforts.
We also thank Profs. Xinmin Zhang and Jianxiong Wang for helpful
discussions.
This work is supported in part by the National Natural Science Foundation
of China under contracts Nos. 19991480, 10175060 and the Chinese Academy
of Sciences under contract No. KJ 95T-03, the 100 Talents Program of CAS
under Contract Nos. U-24, U-25, and the Knowledge Innovation Project of
CAS under Contract Nos. U-602, U-34 (IHEP); by the National Natural Science
Foundation of China under Contract No.10175060(USTC); and
by the Department of Energy under Contract Nos.
DE-FG03-93ER40788 (Colorado State University),
DE-AC03-76SF00515 (SLAC), DE-FG03-94ER40833 (U Hawaii),
DE-FG03-95ER40925 (UT Dallas).

\end{document}